\documentclass[twocolumn,amssymb, nobibnotes, floatfix, aps, prb]{revtex4-2}
\usepackage{hyperref}
\usepackage[utf8]{inputenc}
\usepackage[english]{babel}
\usepackage{amssymb}
\usepackage{amsmath}
\usepackage{graphicx}
\usepackage{color}
\usepackage{physics}
\usepackage{float}
\usepackage{mathtools}
\renewcommand{\vec}[1]{\boldsymbol{#1}}
\begin{document}

%\preprint{}

\title{Lattice Deformation, Low Energy Models and Flat Bands in Twisted Graphene Bilayers}

\author{Niels R. Walet$^1$}
%\email{}
\author{Francisco Guinea$^{1,2}$}
%\email{}
%\thanks{}
\affiliation{$^1$School of Physics and Astronomy, University of Manchester, Manchester, M13 9PY, UK}
\affiliation{$^2$Imdea Nanoscience, Faraday 9, 28015 Madrid, Spain}

\date{\today}

\begin{abstract}
Twisted graphene bilayers show a complex electronic structure, further modified by interaction effects. The main features can be obtained from effective models, which make use a few phenomenological parameters. We analyze the influence of effects at the atomic scale, such as interlayer hopping and lattice relaxation, on the electronic bands. We assume that the twist angle and the size of the Moir\'e pattern is fixed, as it is usually the case in experiments. We obtain a strong dependence of the electronic structure on details of the models at the atomic scale. We discuss how to incorporate this dependence on effective models.
\end{abstract}

\pacs{???}
%\keywords{}

\maketitle

{\it Introduction.}
The discovery of superconductivity and insulating behavior in twisted graphene bilayers\cite{Ketal17,Cetal18a,Cetal18b,Yetal19} has opened new perspectives in the study of two dimensional materials. These systems show a rich phenomenology, likely due to the interplay of a complex electronic structure and the effects of electron interactions. The twist angle, $\theta$, between the two layers defines a Moir\'e structure of lattice unit $L_M = d / \bigl(2 \sin ( \theta / 2 )\bigr)$, where $d \approx 2.42\,\text{\AA}$ is the lattice unit of graphene. New phases appear for small angles, $\theta \lesssim 1.1^\circ$, which corresponds to lattice periodicity $L_M \gtrsim 15$ nm. The fact that $L_M$ is so much larger than the graphene lattice spacing $d$ allows for an effective continuum description of the low-energy electronic bands in terms of a few phenomenological parameters\cite{LPN07,BM11}. These parameters encode averages of the interlayer hopping between carbon $p_z$ orbitals located in the different sublattices of the two layers. The models used so far describe the interlayer hopping by the contribution of three harmonics in the Moir\'e unit cell, and it requires only two parameters, which describe $AA$ and $AB$ hoppings.

\begin{figure}
\begin{center}
\includegraphics[width=0.8\columnwidth]{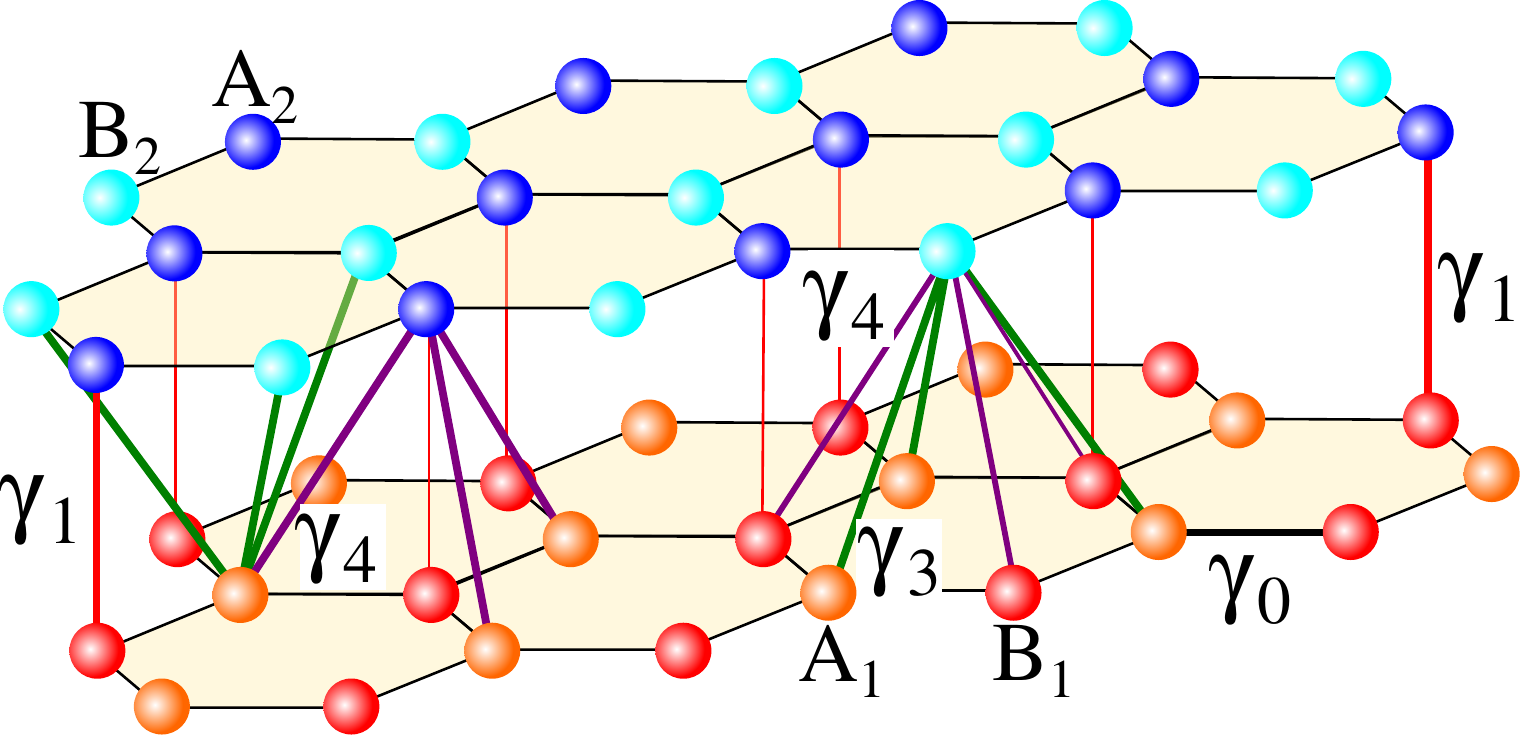} 
\caption{Sketch of the interlayer hoppings in a graphene bilayer.}
\label{fig:sketch}
\end{center}
\end{figure}

These two parameters used in the continuum models, $\{ u_{AA} , u_{AB} \}$, depend on details of the atomic arrangements, and on the way electrons propagate from one layer to the other at the atomic scale. The simplest derivation of the value of these parameters suggest that $u_{AA} = u_{AB} = \gamma_1 / 3$, where $\gamma_1 \approx 0.4$ eV is the hopping between carbon $p_z$ orbitals which are nearest neighbors in different layers in graphite. More sophisticated treatments allow for the inclusion of other interlayer hopping processes which are known to be present in graphite\cite{M10}, like the nearest-neighbor terms $\gamma_3$ and $\gamma_4$, in standard models for graphite\cite{M57,SW58,mcclure_theory_1960,DD02,NGPNG09}, see Fig.~\ref{fig:sketch}. The effect of the atomic displacements caused by the atomic relaxation due to the layer misalignment need also be added, see, e.g., Ref.~\cite{KYKOKF18}. Both the fact that $\gamma_3 \ne \gamma_4$ and lattice relaxation  lead to $g_1 \ne g_2$. For sufficiently low angles, we expect lattice relaxation to be large, as it is energetically favorable to have large areas with $AB$ and $BA$ stacking, while the $AA$ regions will shrink. In this regime, the interlayer hopping is likely to require a description with many harmonics.

In the following, we analyze the influence of the lattice relaxation, and of the dependence of the interlayer hopping on local environment, on the low energy bands of twisted graphene. The lattice relaxation is analyzed using different classical potential models for the interatomic forces. The electronic structure is studied combining different lattice relaxations and different tight binding models for the interlayer hopping, at a fixed angle $\theta = 1.05^\circ$. This angle is close to the values reported experimentally\cite{Cetal18b,Cetal18a}, and defines a commensurate lattice at the atomic level, with a superlattice unit vector $\vec{b}_1 = 32 \vec{ a}_1  + 31 \vec{ a}_2$, where $\vec{ a}_1$ and $\vec{ a}_2$ are the lattice unit vectors of graphene. A few DFT studies are available in the literature\cite{CMFCLK17,SWSLFB18,Letal19}, although it does not (yet) seem feasible to carry out calculations at the size required to deal with small twist angles. Note, finally, that experiments determine the twist angles mostly from measuring the electron density required to fill the bands in the Moir\'e superlattice. The angles observed in this way need not coincide with the theoretically defined ``magic angles'' where the Fermi velocity at the $K$ and $K'$ points in the superlattice Brillouin Zone vanishes\cite{BM11}. Also,  it may make sense to use other definitions of the magic angles, such as those which lead to the narrowest bands, or to the largest gaps between the lowest states and the next ones.

We first consider the role of lattice relaxation, focusing mostly on in-plane relaxation, which is by far the dominant effect. Out of plane relaxation is likely suppressed in encapsulated samples, and, in any case, it leads to weaker effects than in plane relaxation, especially at small angles. Then, we analyze models for the hopping of electrons between different layers. The relation between these tight binding calculations of the electronic structure and continuum approximations is considered next.

\begin{figure}
\begin{center}
\includegraphics[width=\columnwidth]{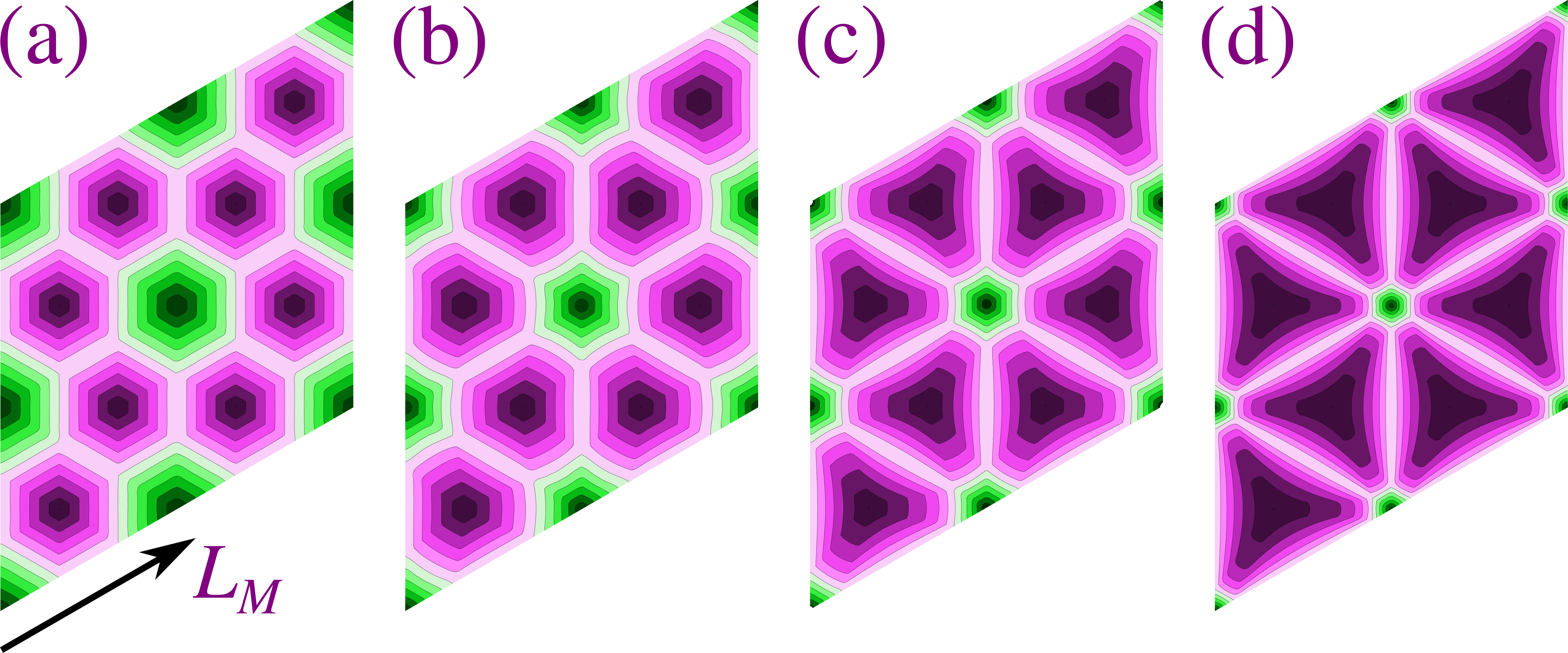} 
\caption{In plane lattice relaxation in twisted bilayer graphene, for a commensurate Moir\'e structure with unit vector ${\bf b} = 32 {\bf a}_1 + 31 {\bf a}_2$, where ${\bf a}_1$ and  ${\bf a}_2$ are unit vectors of the graphene lattice.  The twist angle is $\theta = 1.05^\circ$. a) No relaxation. b) Relaxation using the parameters in Ref.~\cite{nam_lattice_2017}. c) LCBOP+KC force model. d) AIREBO-M+ILP. The color map ranges from dark green for $AA$ alignment to purple for the $AB$ case. White indicates equal $AA$ and $AB$ alignment, and the scale is the same in all figures. In the for cases four unit cells are shown, with $AA$ registry at the corners and midpoint.}
\label{fig:relaxation}
\end{center}
\end{figure}

{\it Lattice relaxation.}
We analyze the lattice relaxation using potential models which reproduce well basic properties of graphene layers and graphite. The in-plane interactions between carbon atoms are described using the AIREBO-M and the LCBOP-I models, which both give the correct nearest neighbor C-C distance\cite{stuart_reactive_2000,los_intrinsic_2003,los_improved_2005,van_wijk_moire_2014,oconnor_airebo-m:_2015}. We have used two forms of interlayer potential: the Kolmogorov-Crespi (KC) potential \cite{kolmogorov_registry-dependent_2005} and the interlayer potential (ILP)\cite{leven_interlayer_2016}. The results seem to be dominated by effects from the interlayer potential, and we have largely concentrated on wto models: LCBOP-I+KC (in short LKC) and AIREBO-M+ILP (which we label as AILP).
We have compared these calculations with the semi-analytical scheme proposed in Ref.~\cite{nam_lattice_2017}, see Ref.~\cite{Long} for details. Some results are shown in Fig.~\ref{fig:relaxation}. We note that the results for the AILP and LKC models that best describe the properties of graphene show a very significant relaxation, which substantially increases the $AB$ and $BA$ aligned regions, and minimizes  the $AA$ regions. One of the way to analyse the correctness of such results is to look at the interface solitons in graphene bilayers \cite{alden_strain_2013}.  We find that both potentials give reasonable results, with a slight preference for the LKC results, which give slightly wider interface solitons in agreement with experiment. The analytical model underestimates the deformation to a great extent.

\begin{figure*}
\includegraphics[width=\textwidth]{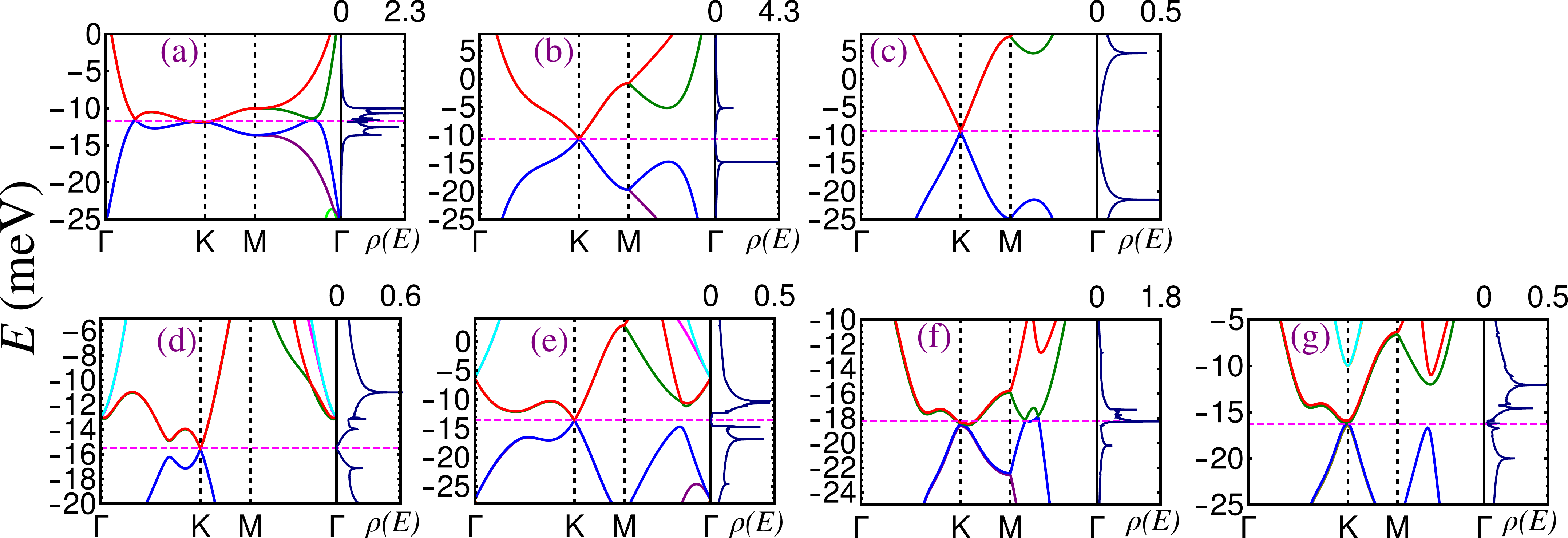} 
\caption{Band structure and density of states of a Moir\'e commensurate superlattice of lattice parameter $\vec{ b}_1 = 32 \vec{ a}_1 + 31  \vec{ a}_2$. The twist angle is $\theta \approx 1.05^\circ$. All of these figures have a constant nearest-neighbor in-layer coupling.
(a)  Undeformed lattice with an exponential Koster-Slater  inter-layer coupling;
(b) LKC deformed lattice with the Koster-Slater coupling;
(c) AILP deformed lattice with the Koster-Slater coupling;
(d) LKC deformed lattice without vertical corrugation with the  screened-1 coupling;
(e) AILP deformed lattice  without vertical corrugation with the screened-1 coupling;
(f) LKC deformed lattice without vertical corrugation with the  screened-2 coupling;
(g) AILP deformed lattice  without vertical corrugation with the screened-2 coupling.}
\label{fig:bands}
\end{figure*}

{\it Tight binding electronic structure.}
We describe the properties of each layer in terms of a single tight binding parameter, which describes the hopping between nearest neighbor $p_z$ orbitals, $\gamma_0 = 2.7$ eV. The interlayer hopping parameters are approximated in two ways: i) couplings between $p$ orbitals at different sites using the Koster-Slater parametrization\cite{SK54} for the relative orientation, and a simple exponential for the distance dependence, with adjustable decay length, $r_0$, and ii) parameters which depend not only on the symmetry of the orbitals and on their distance, but also on the environment near these orbitals. Models in class ii) are required in order to distinguish the interlayer hopping parameters $\gamma_3$ and $\gamma_4$ in graphite and aligned bilayers\cite{M57,SW58,DD02,NGPNG09}, as both  involve carbon atoms at the same distance, and are equal in the Koster-Slater scheme. We have studied the dependence of the electronic structure on the decay length $r_0$ in models of class i), and we have used two choices for the parametrization of the dependence on the local environment in models of class ii), see refs.\cite{sboychakov_electronic_2015,SRRN17}. A comparison of the band structure obtained using a selection of different models, with, and without, lattice relaxation, and with, and without, environment dependent hopping parameters is shown in Fig.~\ref{fig:bands}. A more complete set can be found in Ref.~\cite{Long}. We notice that we have a relatively simple spectrum for the Koster-Slater hopping parameters, Fig.~\ref{fig:bands}a. We have narrow bands, but the angle is smaller than the magic angle, or the interaction is slightly too strong. If we apply the environment-dependent hoppings without lattice deformation, we see that the band-splittings increase, and the bands are not really flat,  Fig.~\ref{fig:bands}b/c. With lattice deformation, either in-plane or fully three-dimensional, we see the appearance of flat narrow structures, with substantial density of state near the Fermi surface. The structure of the bands is rather different, which should also be reflected in the associated electron densities.  

\begin{figure}
\begin{center}
\includegraphics[width=\columnwidth]{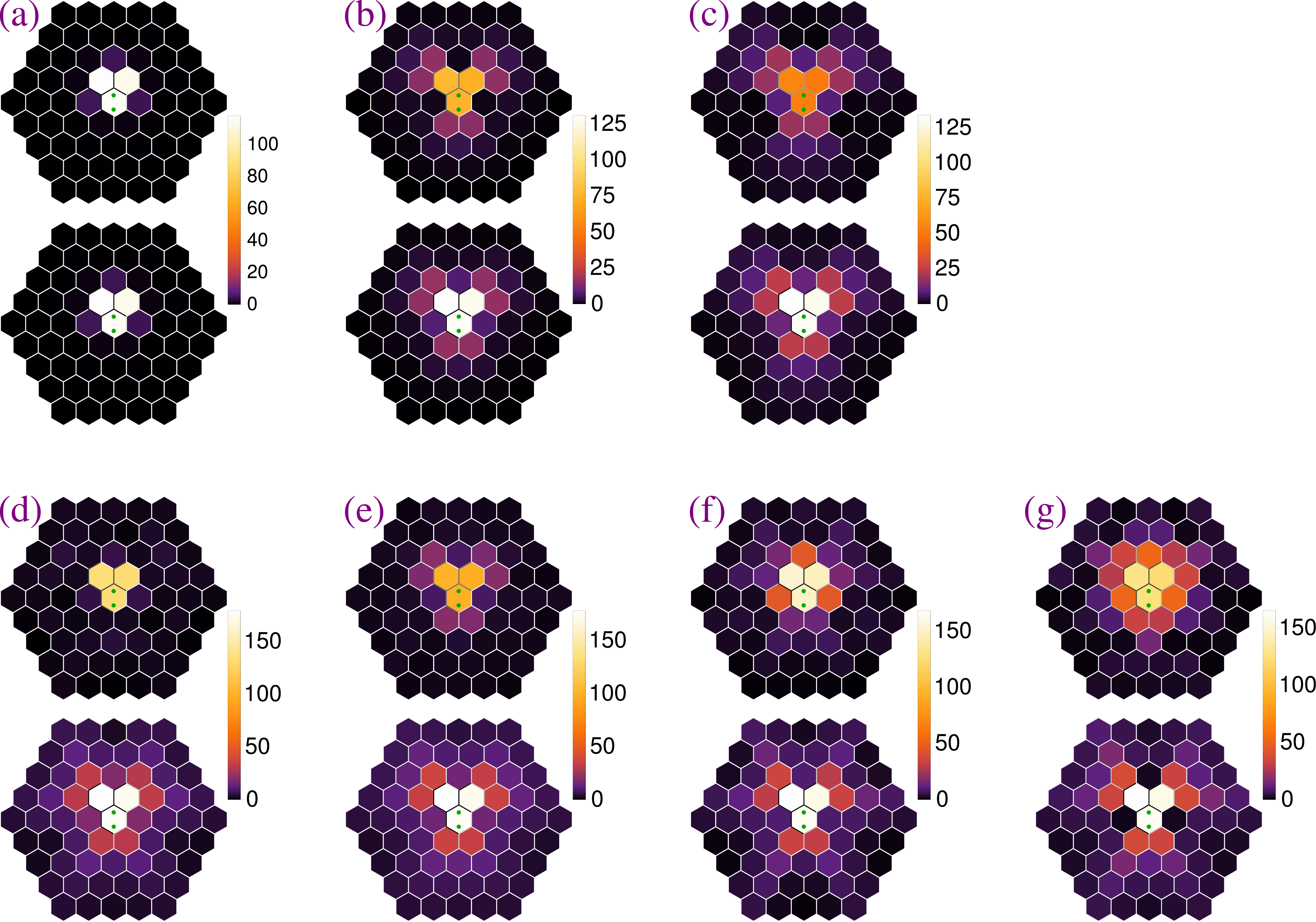} 
\caption{Magnitude of the harmonics of the interlayer hopping elements extracted from the tight binding calculations. Each hexagon stands for a superlattice vector $\bf G$, and the color scale shows the absolute value of each matrix element, in units of meV as labelled in the color bars. The plots corresponds to the bands structures in Fig.~\ref{fig:bands} with the same label.}
\label{fig:harmonics}
\end{center}
\end{figure}

{\it Continuum approximations.}
We analyze here which variations of the continuum approximation best approximate the electronic bands obtained with tight binding models. We expand the $AA$, $AB$ and $BA$ interlayer couplings in superlattice harmonics. In order to do so, we first define the matrix elements
\begin{align}
H ( \vec{ k}, \vec{ G}_m , \vec{ G}_n )&=\prescript{}{1}\langle \vec{k} +\vec{ G}_{\vec m} , s_1 | H(\vec{ k}) | \vec{\bf k}  +\vec{G}_{\vec n} , s'_2 \rangle_2\,.
\label{matrixelement}
\end{align}
where $| H(\vec{ k}) |$ is the tight binding hamiltonian for momentum $\vec{k}$ (defined in the superlattice Brillouin Zone), lndices $1$ and $2$ label the two layers, and $s$ and $s'$ label the sublattices in each layer. The continuum approximation\cite{LPN07,M11,BM11} assumes that the interlayer hopping, which in real space can be written as $V_{s,s'} ( \vec{r}_1 , \vec{r}_2 )$, can be approximated by a local function, \[V_{s,s'} ( \vec{r}_1 , \vec{r}_2 ) \approx V_{s,s'}  ( \vec{r}_1 - \vec{r}_2 )~ .\]  The approximation becomes exact as the twist angle goes to zero, and the size of the Moir\'e unit cell goes to infinity, as the interlayer hopping varies slowly in space. This approximation implies that the matrix element $H ( \vec{ k}, \vec{ G}_m , \vec{ G}_n )$ in Eq.~(\ref{matrixelement})  depends only on the momentum difference, $\vec{G}_m - \vec{G}_n$. The dependence of the matrix elements on $\vec{G}_m - \vec{G}_n$, extracted from different tight binding models, is shown in Fig.~\ref{fig:harmonics}.

\begin{figure}
\begin{center}
\includegraphics[width=\columnwidth]{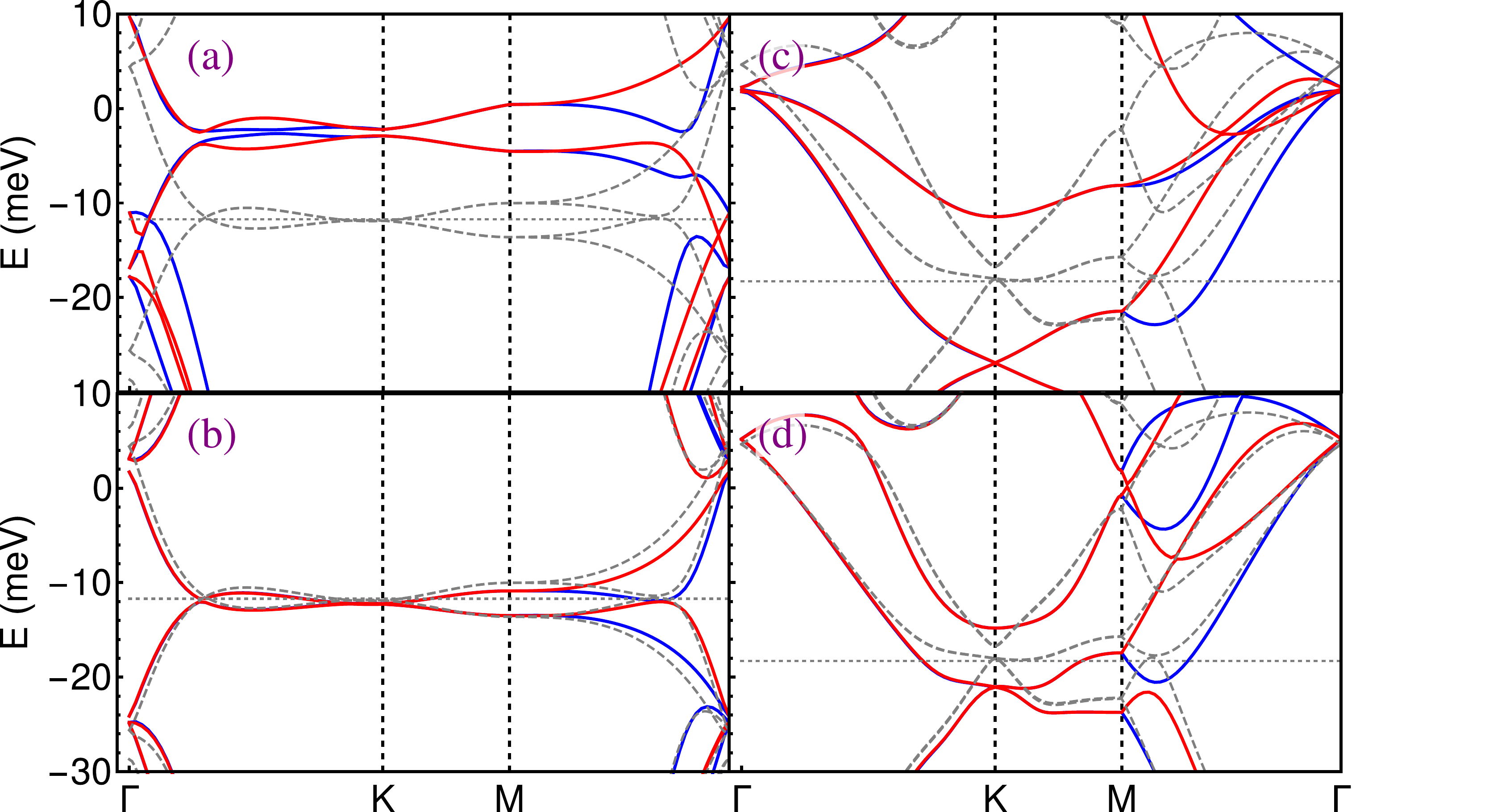} 
\caption{All: Dashed gray curves: tight binding bands.
(a) and (b): calculated for an undeformed lattice and a parametrization of the hoppings with a simple Koster-Slater dependence on distance. (a) Blue and red curves: results from the continuum model where the interlayer hoppings are expanded using three harmonics as in Refs.~\cite{LPN07,BM11}. (b)  as in (a) with the full in-layer tight binding kinetic energy, and a linear $k$ dependence of the interlayer hopping elements. (c) and (d): results for a deformed lattice, using the LKC model, and environment dependent hopping parameters. (c)continuum bands calculated using the same assumptions as in (b). (d) the continuum model as in (c) including the top 48 harmonics, see Fig.~\ref{fig:harmonics}.}
\label{fig:contbands}
\end{center}
\end{figure}

A comparison between tight binding calculations and results from the continuum model with different number of Fourier components of the interlayer hopping, and different approximations for the in layer kinetic energy is shown in Fig.~\ref{fig:contbands}.

For the case studied here, $\theta = 1.05^\circ$, the dependence on the matrix elements in Eq.~(\ref{matrixelement}) on the average momentum, $\vec{k} + ( \vec{G}_m + \vec{G}_n ) / 2$, is not completely negligible. In order to reproduce precisely the tight binding results using a continuum expansion of the interlayer hopping in superlattice harmonics, we need  to both include the dependence of the matrix element, Eq.~(\ref{matrixelement}), on both $\vec{G}_m - \vec{G_n}$ and $\vec{k}$, ii) to take into account lattice corrections (trigonal deformations) to the Dirac equation within each layer, and iii) to include more than three harmonics in the more complex cases. A detailed comparison between tight binding and continuum calculations is made in Fig.~\ref{fig:contbands}.  We see that for the bands in Fig.~\ref{fig:bands}a The standard continuum model works reasonably well--the small symmetry breaking in Fig.~\ref{fig:contbands}a can be fixed, but that the continuum approximation can not describe the particle-hole symmetry breaking. If we add the more complete momentum dependence given above, the results are reproduced almost verbatim; adding a few more harmonics gives perfect results \cite{Long}. For the much more complex shown in  Fig.~\ref{fig:contbands}c/d we see that the expansion converges more slowly, but that a good continuum model can be constructed.

{\it Conclusions.}
We have analyzed lattice relaxation in twisted graphene layers, using different atomic force models. We find that the models which fit better the properties of of aligned bilayers and graphite lead to a significant relaxation, with large regions of $AB$ and $BA$ stacking, while the $AA$ regions are reduced. The twist angle used in these calculations is $\theta = 1.05^\circ$.

The relaxed positions of the atoms are used as input for the calculation of the electronic structure, calculated using tight binding models. Different parametrizations of the couplings are used: i) hoppings between orbitals in different layers which combine a form factor which reflects the symmetry of $p$ orbitals, and a simple exponential dependence on distance, and ii) hoppings that depend on the distance {\it and} the local environment of the two orbitals involved in the process. Models of type ii) reproduce the difference between the hoppings $\gamma_3$ and $\gamma_4$ needed to describe aligned bilayers and graphite. For a fixed twist angle, $\theta \approx 1.05^\circ$, the low energy bands show a significant dependence on both the range of the interaction and whether the hopping parameters depend solely on interatomic distances, or they also include other features of the environment. To some extent, the results can be interpreted as a parameter dependent shift of the ``magic angle'', where the low energy bands are narrowest. When the choice of parameters is such that the magic angle is greater than $1.05^\circ$, we find new band crossings and Dirac points\cite{HLSCB19}.

We have analyzed the minimal continuum models required to approximate the electronic bands obtained from tight binding calculations defined at the atomic scale. The complexity of the continuum models depends significantly on the range of the hoppings, and on whether they depend significantly on the local environment. Isotropic couplings which do not decay too abruptly with distance are reasonably described with the standard model based on an expansion with three harmonics of the interlayer hoppings. A continuum description is possible for all tight binding models considered, although more than three harmonics are required in some cases, especially when the hoppings depend on the local environment.

We have compared results from various models,  both for the interatomic forces and for the electronic hoppings, using the same twist angle, $\theta = 1.05^\circ$. This choice is motivated by the fact that the value of the twist angle is the magnitude most accessible experimentally. It is yet unclear how precisely the experimentally studied twist angles correspond to the theoretical definition of magic angles. The dependence found here of the electronic properties on the choice of parameters suggests that the observed tendency towards broken symmetry phases must be quite robust. The appearance of superconductivity and insulating behavior in twisted graphene bilayers is likely to arise from rather general properties of the models.

{\it Acknowledgements.}
FG  was  supported  by
the European Commission under the Graphene Flagship,
contract CNECTICT-604391; NRW is supported by  UK STFC under grant  ST/P004423/1.
\bibliography{moire_v3}
\end{document}